\documentclass[%
reprint,
longbibliography,
amsmath,
amssymb,
aps,
prapplied,
floatfix,
twocolumn,
]{revtex4-1} 

\usepackage{graphicx}
\usepackage{siunitx}
\usepackage{amsmath}

\usepackage{hyperref}

\hypersetup{
    colorlinks=true,       
    linkcolor=blue,          
    citecolor=blue,        
    filecolor=blue,      
    urlcolor=blue           
}

\graphicspath{{./}{./figs/}}

\begin{document}

\title{Single-beam Zeeman slower and magneto-optical trap using a nanofabricated grating}

\author{D. S. Barker}
\email[]{daniel.barker@nist.gov}
\author{E. B. Norrgard}
\author{N. N. Klimov}
\author{J. A. Fedchak}
\author{J. Scherschligt}
\author{S. Eckel}
\email[]{stephen.eckel@nist.gov}
\affiliation{Sensor Science Division, National Institute of Standards and Technology,
Gaithersburg, MD 20899, USA}


\begin{abstract}
We demonstrate a compact (\(0.25~\si{\liter}\)) system for laser cooling and trapping atoms from a heated dispenser source.
Our system uses a nanofabricated diffraction grating to generate a magneto-optical trap (MOT) using a single input laser beam.
An aperture in the grating allows atoms from the dispenser to be loaded from behind the chip, increasing the interaction distance of atoms with the cooling light.
To take full advantage of this increased distance, we extend the magnetic field gradient of the MOT to create a Zeeman slower.
The MOT traps approximately \(10^6\) \(^7\)Li atoms emitted from an effusive source with loading rates greater than \(10^6~\si{\per\second}\).
Our design is portable to a variety of atomic and molecular species and could be a principal component of miniaturized cold-atom-based technologies.
\end{abstract}



\maketitle

\section{Introduction}

Miniaturized cold-atom systems may form the basis of a host of emerging quantum technologies, from quantum repeaters~\cite{Kuzmich2003} to clocks~\cite{Koller2017a}.
Such miniaturized systems will likely employ a magneto-optical trap (MOT) for initial cooling and trapping of atoms.
Conventional MOTs confine an atomic gas near the center of a quadrupole magnetic field in the overlap region of three pairs of counterpropagating laser beams~\cite{Raab1987}.
Due to the number of laser beams, MOTs typically have expansive optical layouts with a large number of mechanical degrees of freedom.
Even mobile experiments that incorporate MOTs have a size on the order of \(1~\si{\meter}\)~\cite{Koller2017a, Hauth2013, Becker2018, Liu2018}.
To fully realize the potential of cold-atom-based quantum technologies beyond the laboratory environment, the size and robustness of MOTs need to be improved.

\begin{figure}[t!]
  \center 
  \includegraphics[width=\columnwidth]{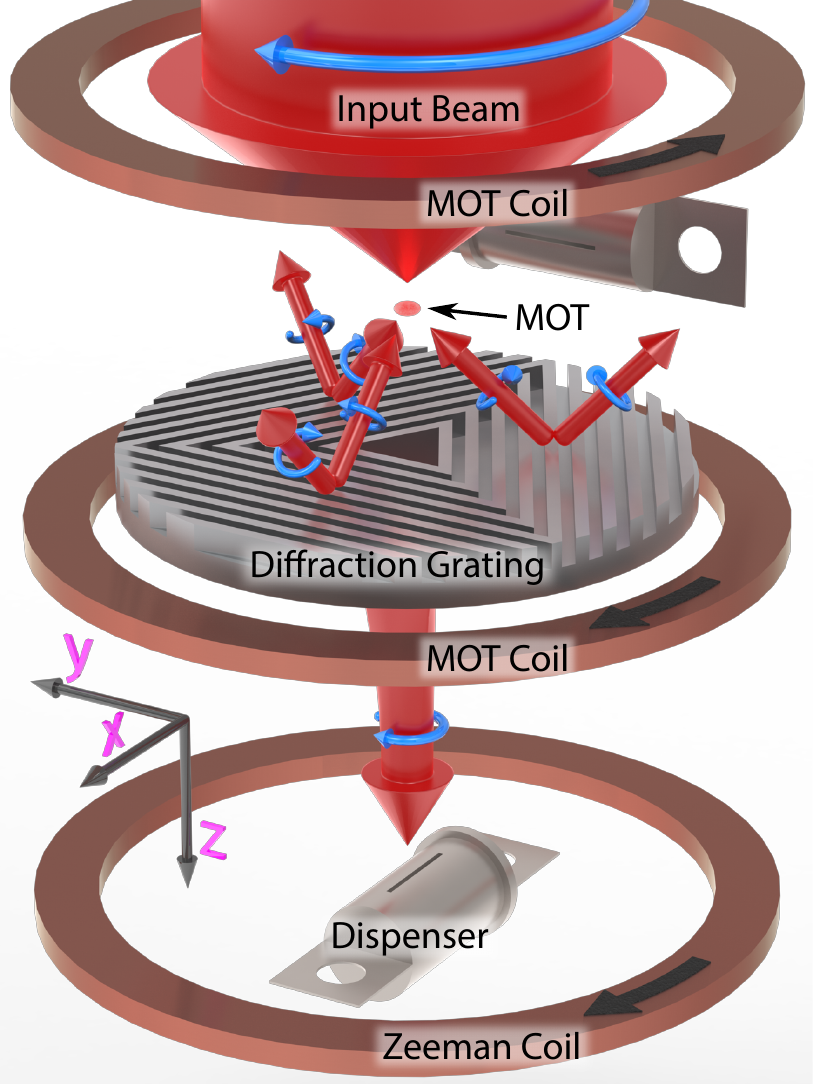}
  \caption{\label{fig:sketch}
  An illustration of the experimental apparatus (not to scale).
  The red arrows depict the input and diffracted laser beams.
  The blue arrows wrapping each laser beam denote its circular polarization.
  The copper rings represent electromagnets and the attached black arrows show the direction of current flow.
  The two MOT coils form a quadrupole magnetic field, while the Zeeman coil increases the magnetic field behind the diffraction grating.
  Lithium atoms that leave the heated dispenser are slowed by the single laser beam behind the grating and then captured by the MOT.\@
  The second dispenser in the background illustrates a hypothetical side-loading configuration, which we compare to our experimental results.
  The axes of ensuing figures refer to the coordinate system shown here.
  The gravitational force is antiparallel to the \(z\) axis.
  }
\end{figure}

Previous research on MOT miniaturization has focused on elements that can be trapped from a room-temperature background vapor, namely Cs or Rb.
However, many other elements can be laser cooled and each have advantages for various quantum technologies.
For example, Sr~\cite{Campbell2017} or Yb~\cite{Hinkley2013} can be used as a highly accurate clock.
Lithium, due to its low mass, has been identified as a possible sensor atom for primary vacuum gauges~\cite{Scherschligt2017, Eckel2018} and, given its large recoil energy, could find use in cold-atom gravity gradiometers~\cite{Sorrentino2010}.
Most atoms, including Li, do not have an appreciable vapor pressure at room temperature and thus are typically loaded from a heated dispenser.
Here, we present the design of a compact laser cooling and trapping apparatus for Li that integrates a MOT with a Zeeman slower and requires only a single input laser beam.
Figure~\ref{fig:sketch} shows the essential features of our apparatus.
We anticipate that our design can be adapted to other elements and possibly molecules.

The two main approaches to MOT miniaturization are based on early experiments using pyramidal retroreflectors~\cite{Lee1996} or tetrahedral laser beam arrangements~\cite{Shimizu1991}.
These pyramidal and tetrahedral MOT configurations allow the formation of a MOT using a single external laser beam and a compound reflective optic~\cite{Lee1996, Vangeleyn2009}.
In the tetrahedral geometry, the compound optic can be fully planarized by replacing the reflectors with diffraction gratings (see Fig.~\ref{fig:sketch})~\cite{Vangeleyn2010}.
Both grating and pyramidal MOTs have been demonstrated to trap large numbers of atoms~\cite{Trupke2006, Bodart2010, Nshii2013, Imhof2017, Wu2017} and cool them below the Doppler limit~\cite{Lett1989, Bodart2010, Nshii2013, Lee2013a, Wu2017}.
Pyramidal MOTs have been made into single-beam atom interferometers~\cite{Bodart2010, Wu2017} and are being developed into compact atomic clocks~\cite{Xu2008a, Scherer2014}.
Grating MOTs have found use as magnetometers~\cite{McGilligan2017} and electron beam sources~\cite{Franssen2019}.
The optics for both MOT types are amenable to nanofabrication~\cite{Trupke2006, Pollock2009a, Pollock2011, Nshii2013, McGilligan2015, McGilligan2016, Cotter2016}.
Nanofabricated pyramidal MOTs are inferior to grating MOTs in two key areas.
First, grating MOTs form above the nanofabricated grating chip, making the laser-cooled atoms easier to manipulate and detect.
Second, the fabricated optics of a grating MOT are planar, making grating MOTs fully compatible with atom chips~\cite{Fortagh2007, Keil2016} and photonics~\cite{Vetsch2010}.

The atom loading rate \(R\) of a MOT depends strongly on the capture velocity \(v_c\).
Namely, \(R \propto {(v_c/v_p)}^4\), where the constant of proportionality depends on the total flux from the source and \(v_p=\sqrt{2 k_B T/m}\) is the most probable thermal velocity of particles with temperature \(T\) and mass \(m\) (\(k_B\) is Boltzmann's constant)~\cite{Lindquist1992}.
While difficult to calculate {\it a priori}, a reasonable upper limit on \(v_c\) is given by the maximum atomic velocity that can be stopped in a distance \(d_s\) (typically a MOT laser beam radius \(r_b\)) by the radiation force, i.e., \(v_c < \sqrt{d_s \hbar k \Gamma/m}\), where \(k=2\pi/\lambda\) is the wavenumber of the cooling light with wavelength \(\lambda\), \(\Gamma\) is the decay rate of the excited state, and \(\hbar\) is the reduced Planck constant. 
The corresponding figure of merit for \(R\) is then \({(d_s \hbar k \Gamma/2 k_B T)}^2\), assuming the same source output flux.
Lithium's figure of merit is among the worst of all laser-coolable atoms, with its red cooling wavelength of \(\lambda_{\text{Li}}\approx 671~\si{\nano\meter}\), its linewidth of \(\Gamma_{\text{Li}}\approx 2\pi\times6~\si{\mega\hertz}\), and its typical source operating temperature \(T\approx 700~\si{\kelvin}\).

Lithium's poor figure of merit is worsened when loading a grating MOT directly from a dispenser.
For simplicity, the dispenser could be placed to the side to avoid blocking laser beams (see Fig.~\ref{fig:sketch}).
This placement results in \(d_s\lesssim \sqrt{2} r_b\) for a conventional six-beam MOT, but only \(d_s\lesssim r_b/2\) for a grating MOT.\@
Moreover, a dispenser placed to the side of a grating MOT will tend to deposit metal on the grating, gradually reducing its performance.

Another important MOT performance metric is the steady-state atom number \(N_{S} = R\tau\), where \(\tau\) is the trap lifetime. 
To achieve the same output flux, different elements require different source temperatures, \(T\).
Higher-temperature sources outgas undesired species at a rate that is exponential in \(T\).
These undesired species can collide with trapped atoms, reducing \(\tau\). 

We exploit the natural integrability of the tetrahedral MOT configuration with a Zeeman slower by etching an aperture in the grating and loading the atoms from behind~\cite{Shimizu1991, Lin1991}.
Light passing through the aperture in the chip can interact with the counterpropagating atoms for a greater distance, increasing \(d_s\) (see Fig.~\ref{fig:sketch}).
By tailoring the magnetic field behind the chip, we make a Zeeman slower to capture atoms with higher initial velocity.
The aperture can also serve as a gas flow limiter, allowing for differential pumping that mitigates the effects of dispenser outgassing.

\section{Description of the trap}

\begin{figure}
    \center 
    \includegraphics[width=\columnwidth]{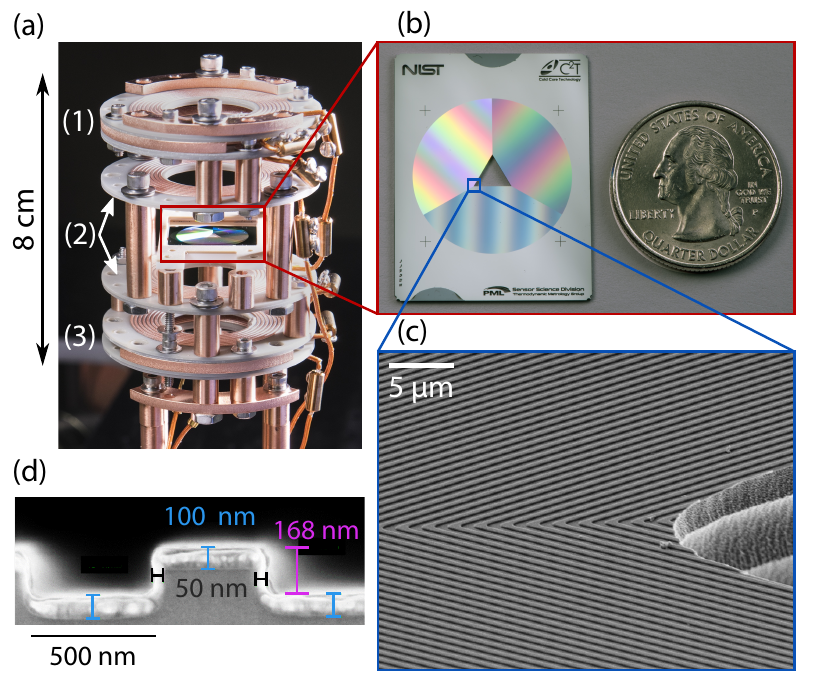}
    \caption{\label{fig:device} The cooling and trapping apparatus for Li.
    (a) A picture of the full device including (1) compensation coils, (2) MOT coils, and (3) Zeeman coils.
    The atom dispenser is concealed by the Zeeman coils, but its approximate position is shown in Fig.~\ref{fig:sketch}.
    (b) The \(27~\si{\milli\meter}\) by \(35~\si{\milli\meter}\) diffraction grating chip with a U.S. quarter for scale.
    (c) A scanning electron microscope (SEM) image of the diffraction grating near a vertex of the triangular aperture.
    (d) An edge-on SEM image with critical dimensions.}
\end{figure}

A nanofabricated silicon diffraction grating chip forms the core of the device, as shown in Fig.~\ref{fig:device}(a). 
The grating chip, fabricated using photolithography at the National Institute of Standards and Technology in the Center for Nanoscale Science and Technology (CNST) NanoFab cleanroom facility~\cite{disclaimer}, consists of three one-dimensional diffraction gratings, which are arranged so that their grooves form concentric equilateral triangles (see Fig.~\ref{fig:device}(b) and Fig.~\ref{fig:device}(c)). 
Each grating has a period \(p=1.00(5)~\si{\micro\meter}\) and a \(500(10)~\si{\nano\meter}\) trench width [see Fig.~\ref{fig:device}(d)] (here, and throughout the paper, parenthetical quantities represent standard uncertainties). 
The gratings are cropped by an outer circle with a diameter of \(22~\si{\milli\meter}\).
The diffraction gratings have a first-order diffraction angle \(\theta_d\approx42\si{\degree}\) at \(\lambda_{\text{Li}}\).
The grating trenches are etched to a depth of \(168(2)~\si{\nano\meter}\) (approximately \(\lambda_{\text{Li}}/4\)), chosen to minimize zero-order reflections.
A \(100(5)~\si{\nano\meter}\) layer of aluminum is deposited on the chip surface.
The aluminum coating thickness is chosen by interpolating the data of Ref.~\cite{McGilligan2016} to yield, at \(\lambda_{\text{Li}}\), the optimum first-order diffraction efficiency of \(33~\si{\percent}\) for a triangular grating MOT;\@ we measure \(37(1)~\si{\percent}\).
Higher-order diffraction is suppressed because \(p<2\lambda_{\text{Li}}\).
For normally incident, circularly polarized light, the normalized Stokes parameters of the first-order diffracted beam are \(Q=0.03(1),\,U=0.13(1),\,V=0.84(1)\).
A triangular aperture, defined by an inscribed circle of radius \(1.5~\si{\milli\meter}\), allows both light and atoms to pass through the chip.

Three sets of electromagnets generate the necessary magnetic fields (see Fig.~\ref{fig:device}(a)). 
Set (2) is an anti-Helmholtz pair that produces the magnetic field gradient needed for the MOT.\@
Set (3) extends the range of the magnetic field beyond the chip and adapts it into the square root profile of a Zeeman slower (see Sec.~\ref{sec:ZS}).
The antisymmetric set (1) prevents the field from set (3) from shifting the MOT axially.
All sets are made from direct bond copper on an aluminum-nitride substrate.

Our Li dispenser is a custom-length, commercially-available vapor source.
It consists of a stainless-steel tube filled with \(15~\si{\milli\gram}\) of unenriched Li.
The dispenser emits atoms through a \(5~\si{\milli\meter}\) by \(0.1~\si{\milli\meter}\) rectangular slit.
Assuming an operating temperature of \(375~\si{\degreeCelsius}\), the dispenser can operate continuously for approximately \(200\)~days before exhausting its Li supply~\cite{Beijerinck1975, Alcock1984}.
Future versions of our trapping system will use a 3D-printed titanium dispenser that can hold more than \(100~\si{\milli\gram}\) of Li~\cite{Norrgard2018}; allowing at least \(500\)~days of continuous operation.

The full \(0.25~\si{\liter}\) assembly is constructed on a standard vacuum flange and inserted into a vacuum chamber pumped by a \(50~\si[per-mode=symbol]{\liter\per\second}\) ion pump.
The vacuum chamber has a base pressure of \(3(1)\times10^{-8}~\si{\pascal}\).
Outgassing from the Li source causes the pressure to increase to approximately \(10^{-6}~\si{\pascal}\).

The single, intensity-stabilized laser beam strikes the grating normally.
It has a \(1/e^2\) radius of \(20(1)~\si{\milli\meter}\); an iris stops the beam to fit the grating.
The center frequency of the laser is detuned relative to the \(^2\)S\(_{1/2}\)(\(F=2\)) to \(^2\)P\(_{3/2}\)(\(F'=3\)) cycling transition, which has a saturation intensity of \(I_\text{sat} \approx 2.54~\si[per-mode=symbol]{\milli\watt\per\square\centi\meter}\).
An electro-optic modulator adds sidebands at approximately \(813~\si{\mega\hertz}\); the \(+1\) sideband is equally detuned from  the \(^2\)S\(_{1/2}\)(\(F=1\)) to
\(^2\)P\(_{3/2}\)(\(F'=2\)) ``repump'' transition.
Because the MOT magnetic field gradient continuously deforms into the Zeeman slower field, an additional slowing laser would not improve atom capture.
The intensities \(I\) reported herein are the carrier intensity at the center of the incident beam.
Fluorescence from the MOT is continuously monitored by a camera along an axis orthogonal to the cooling beam.
The same camera also records absorption images to more accurately measure the number of trapped atoms.

\begin{figure}[t]
    \center 
    \includegraphics{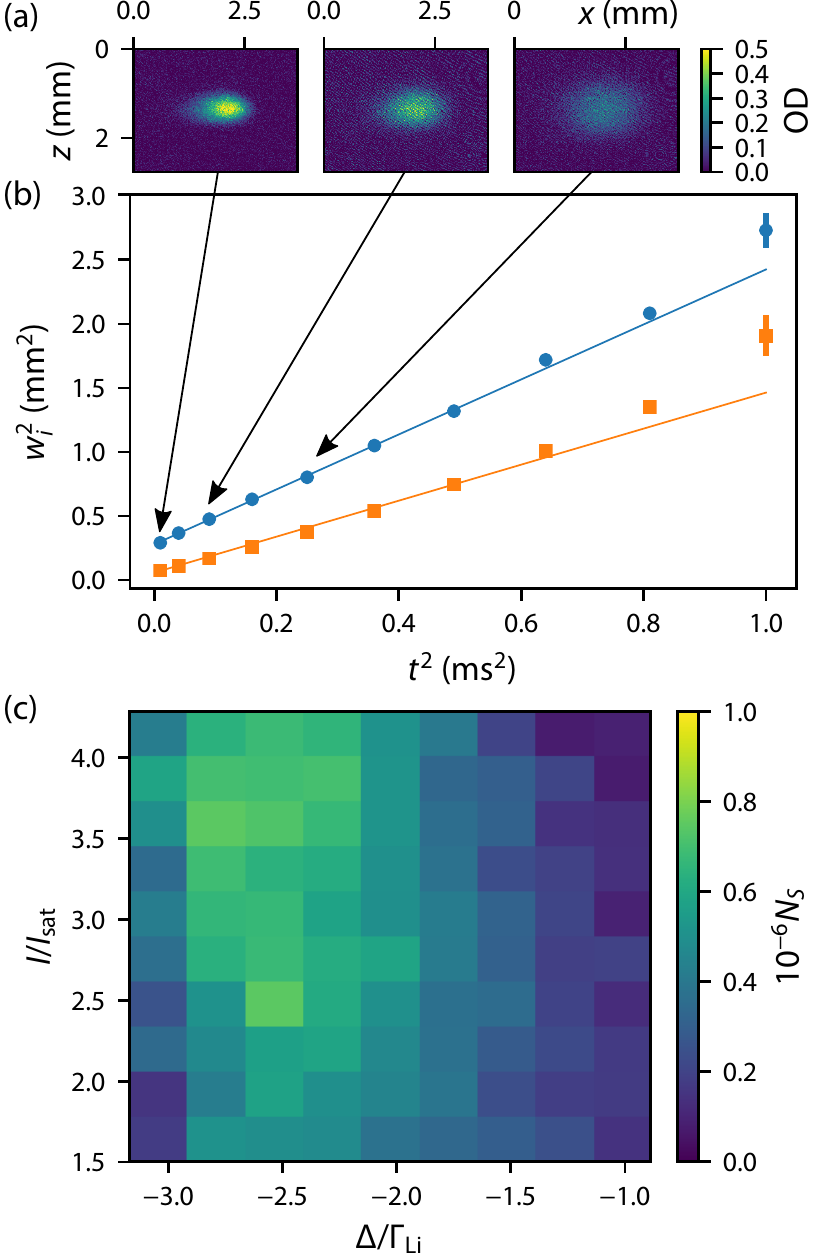}
    \caption{\label{fig:atom_number} The MOT performance.
    (a) Absorption images showing the optical density (OD) of the \(^7\)Li cloud at three different expansion times \(t\).
    (b) The fitted \(x\) (blue) and \(z\) (orange) squared \(1/e\) radii \(w^{2}_{i=x,z}\) of the cloud vs.\ \(t^2\) (most error bars are smaller than the data points).
    (c) The steady-state atom number \(N_{S}\) vs.\ the laser detuning and intensity, in units of \(\Gamma_{\text{Li}}\) and \(I_\text{sat}\), respectively.}
\end{figure}

\section{MOT parameters}

Figure~\ref{fig:atom_number}(a) shows images of the cloud of \(^7\)Li atoms previously trapped in the MOT after various expansion times \(t\). 
The diffraction grating is not visible in the images, as the MOT forms approximately \(4~\si{\milli\meter}\) from the chip.
The fitted \(1/e\) radii of the cloud, shown in Fig.~\ref{fig:atom_number}(b), follow the expected \(w^2(t) = w_0^2 + (2 k_B T_\text{MOT}/m) t^2\), where \(T_\text{MOT}\) is the temperature of the trapped cloud and \(w_0\) is the initial radius. 
The measured temperatures are \(900(50)~\si{\micro\kelvin}\) in the radial direction and \(590(30)~\si{\micro\kelvin}\) in the axial direction.
For the data in Fig.~\ref{fig:atom_number}(a) and Fig.~\ref{fig:atom_number}(b), the laser detuning is \(\Delta/\Gamma_{\text{Li}} = -2.0\), the saturation parameter is \(s_{0} = I/I_\text{sat} = 3.6\), the carrier-to-repump power ratio is about \(3:2\), the magnetic field gradient at the center of the MOT is \(B' = 4.5~\si[per-mode=symbol]{\milli\tesla\per\centi\meter}\), and the peak magnetic field of the Zeeman slower is \(B_\text{max}\approx 12~\si{\milli\tesla}\). 
These trapping parameters and the extracted temperatures are similar to those of ``compressed'' Li MOTs reported in the literature~\cite{Mewes2000, Grier2013, Burchianti2014}.

Figure~\ref{fig:atom_number}(c) shows the equilibrium atom number \(N_{S}\) in the MOT as a function of detuning and intensity of the laser beam. 
The magnetic field gradient, the peak Zeeman slower field, and the carrier-to-repump ratio are the same as in Fig.~\ref{fig:atom_number}(a) and Fig.~\ref{fig:atom_number}(b). 
The captured atom number increases with the laser intensity and begins to saturate at \(I/I_\text{sat}\approx 2.5\).
We find that the maximum atom number occurs near a detuning of \(\Delta/\Gamma_{\text{Li}}\approx -2.5\).
Varying the carrier-to-repump power ratio between \(1:1\) and \(2:1\) does not qualitatively change the results in Fig.~\ref{fig:atom_number}(c) or substantially affect the maximum atom number. 

The simplicity of our setup complicates measurement of the MOT lifetime.
There is no distinct Zeeman slowing laser beam and the Li dispenser takes minutes to turn off, so the MOT always loads atoms during operation.
However, we can shut off the current in the Zeeman and compensation coils (see Fig.~\ref{fig:sketch} and Fig.~\ref{fig:device}(a)) to drastically reduce the MOT loading rate (see Sec.~\ref{sec:ZS}).
The MOT population then exponentially decays to a lower equilibrium atom number.
Fitting the MOT decay curves yields trap lifetimes \(\tau \approx 1~\si{\second}\) for our operating conditions.

\section{Zeeman slower performance}\label{sec:ZS}

To quantitatively understand the loading of the MOT, we calculate the average
force \(\mathbf{f}\) exerted on an atom by the input beam, with wavevector
\(\mathbf{k}_0\), and its reflections, with wavevector \(\mathbf{k}_i\) (\(i=1,2,3\)), using
\begin{equation}
    \label{eq:force}
    \begin{split}
    \mathbf{f} & = \sum_{i=0}^{3}\frac{\hbar \mathbf{k}_i \Gamma}{2}
    \sum_{m_L'=-1}^1\frac{s_i P(m_L',\gamma_i,\epsilon_i)}
    {1+s_\text{total}+4\delta_{i}^2/\Gamma^2}, \\
    \delta_{i} & = \Delta - \mathbf{k}_i\cdot\mathbf{v} - m_L' \mu_B B/\hbar
    \end{split}
\end{equation}
where we have assumed an S-to-P transition (i.e., ignoring fine and hyperfine
structure)~\cite{Vangeleyn2009}.
Here, \(s_i = I_i/I_\text{sat}\) is the saturation parameter for beam
\(i\) (with intensity \(I_i\)), \(s_\text{total}=\sum_{\,i}s_i\), \(\Delta\) is the detuning, \(\mathbf{v}\) is the atom's velocity, \(\mu_B\) is the Bohr magneton, \(m_L'\) is the projection of the excited state's orbital angular momentum onto the magnetic field, and \(\gamma_i\) is the angle between the magnetic field \(\mathbf{B}\) and wavevector \(\mathbf{k}_i\). 
The polarization of the beam \(i\) is denoted by \(\epsilon_i=\pm 1\), where \(+1\) (\(-1\)) represents right-handed (left-handed) circular polarization.
\(P\) is a Wigner \(d\)-matrix that determines the transition probabilities to excited state \(m_L'\) and is given by \(P(m_L'=-1, \gamma_i, \epsilon_i=\pm 1) = {(1 \mp \cos\gamma_i)}^2/4\), \(P(m_L'=0, \gamma_i,\epsilon_i=\pm 1) = \sin^2\gamma_i/2\), and \(P(m_L'=+1, \gamma_i, \epsilon_i=\pm 1) = {(1 \pm \cos\gamma_i)}^2/4\). 

\begin{figure}[t]
    \center 
    \includegraphics[width=\columnwidth]{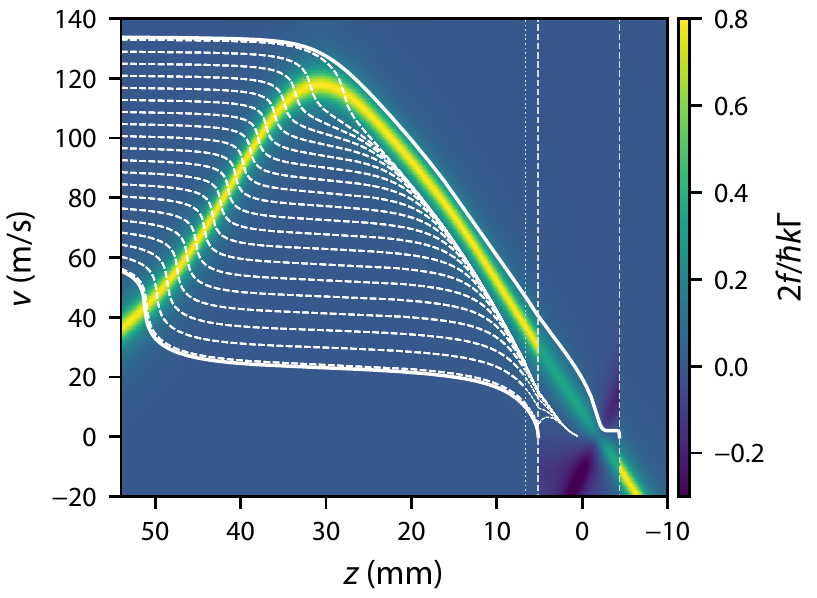}
    \caption{\label{fig:slower} Zeeman slower forces on the atoms.
    The color plot shows the axial force on a \(^7\)Li atom vs.\ position \(z\) and axial velocity \(v\) for \(B' = 4.5~\si[per-mode=symbol]{\milli\tesla\per\centi\meter}\) and \(B_\text{max} = 12~\si{\milli\tesla}\).
    The vertical dashed lines denote the MOT region; the vertical short-dashed line denotes the chip location.
    The atom source is positioned at the left edge of the plot (\(z=z_s\approx54~\si{\milli\meter}\)).
    The solid white curves show the trajectories of the slowest and fastest captured velocities.
    The dashed white curves show the trajectories for intermediate initial velocities.}
\end{figure}

\begin{figure*}[t]
    \center 
    \includegraphics[width=\textwidth]{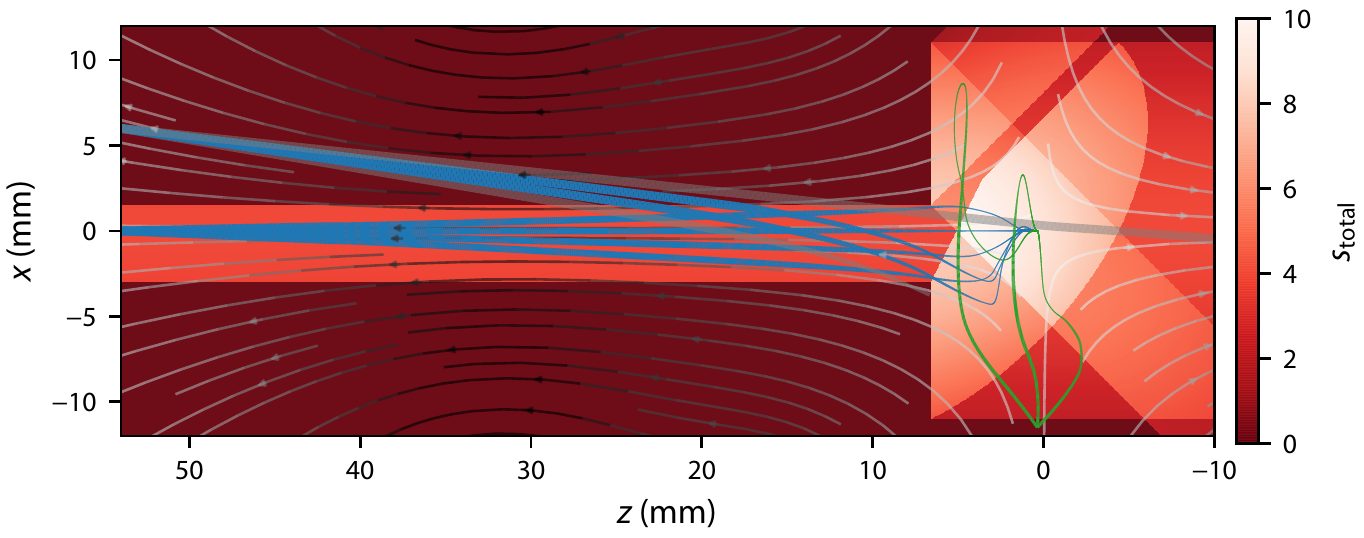}
    \caption{\label{fig:traj} Off-axis trajectories of atoms.
    The color plot shows the total saturation parameter of all lasers, \(s_\text{total}\); the stream lines show magnetic field lines with a larger magnetic field magnitude being darker.
    Blue (green) trajectories are trapped trajectories from a source placed behind (to the side of) the chip, starting at \(0.7v_c\) (\(0.25v_c\)); gray trajectories are untrapped.
    The width of the trajectory curves indicates the magnitude of the velocity.}
\end{figure*}

The calculated force along the \(z\) axis is shown in Fig.~\ref{fig:slower} for a magnetic field gradient of \(B' = 4.5~\si[per-mode=symbol]{\milli\tesla\per\centi\meter}\) and a maximum magnetic field of \(B_\text{max}\approx 12~\si{\milli\tesla}\). 
For these values of \(B'\) and \(B_\text{max}\), the magnetic field behind the chip closely matches the ideal \(B(z)\propto\sqrt{z}\) Zeeman slower profile.
The input laser beam (see Fig.~\ref{fig:sketch}) is resonant with the cycling transition along the bright yellow curve (i.e., \(kv = -\Delta + \mu_B B/\hbar\)), maximizing the slowing force \(\mathbf{f}\) (see Eq.~\ref{eq:force}).
The force is reduced in the MOT region (the yellow curve darkens to pale green) because the diffracted laser beams increase \(s_\text{total}\) (see Fig.~\ref{fig:traj}).
The dispenser source is located at \(z_s\approx54~\si{\milli\meter}\), beyond the maximum of the magnetic field at \(z_\text{max}\approx 30~\si{\milli\meter}\), and about \(10\) times further from the MOT than the aperture at \(z_a=5~\si{\milli\meter}\) with characteristic radius \(r_a=2~\si{\milli\meter}\).

Behind the aperture (see Fig.~\ref{fig:sketch}), atoms are slowed similarly to an ideal Zeeman slower, where the velocity follows \(v_B(z) = \mu_B B(z)/\hbar k\).
In this case, all initial velocities \(v_0<v_c = \mu_B B_\text{max}/\hbar k\) should be slowed.
Fig.~\ref{fig:slower} shows simulated on-axis \(v(z)\) trajectories. 
Atoms emitted from the source with \(v_0\) receive a slowing impulse as they come into resonance with the slowing laser beam in the region of increasing magnetic field (\(z\gtrsim 30~\si{\milli\meter}\) in Fig.~\ref{fig:slower}), travel along the Zeeman slower at nearly constant velocity, and then fall onto the \(v_B(z)\) curve. 

\begin{figure}
    \center 
    \includegraphics{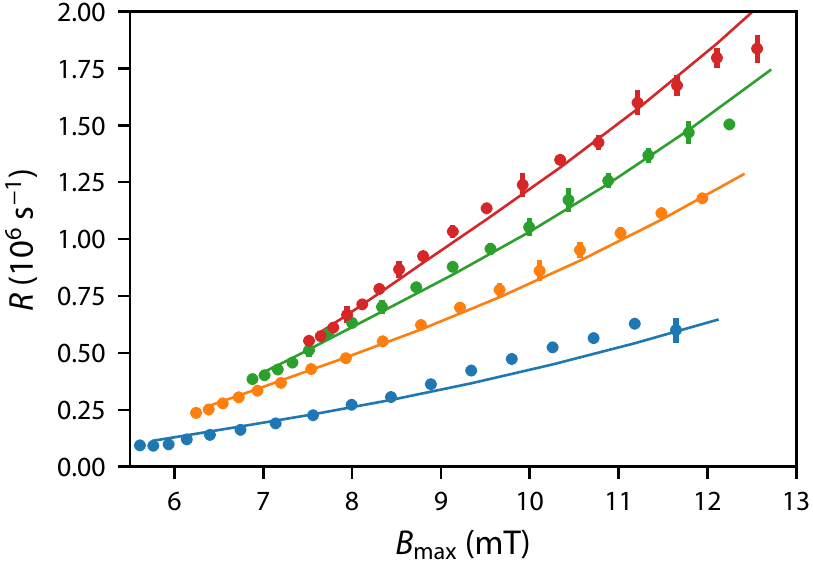}
    \caption{\label{fig:slower_efficacy} The measured loading rate of the MOT with \(\Delta/\Gamma_{\text{Li}} = -2.0\) and \(s_{0}=I/I_\text{sat}=3.9\) at four different magnetic field gradients \(B'\): \(4~\si[per-mode=symbol]{\milli\tesla\per\centi\meter}\) (blue), \(4.5~\si[per-mode=symbol]{\milli\tesla\per\centi\meter}\) (orange), \(5~\si[per-mode=symbol]{\milli\tesla\per\centi\meter}\) (green), \(5.5~\si[per-mode=symbol]{\milli\tesla\per\centi\meter}\) (red).
    The curves are best fits of the model described in the text.}
\end{figure}

We calculate the resulting loading rate by considering an effusive source with surface area \(\mathcal{S}\).
Each area element of the source \(d\mathcal{S}\) emits \(\phi\) atoms per second per unit area per steradian according to a cosine distribution~\cite{Ross1995}. 
Due to the size of the chip aperture and MOT beams, only atoms emitted at angles \(\theta\) (relative to the \(z\) axis) less than the capture angle \(\theta_c\) are captured by the MOT.\@ 
Integrating over the full source surface \(\mathcal{S}\) leads to
\begin{equation}
    \label{eq:loading_rate}
    R = 8 \sqrt{\pi} \int_\mathcal{S} \phi \,d\mathcal{S} \int^{\theta_c}_0 \cos\theta \,d\theta \int_0^{v_c(\theta)} \frac{v_0^3}{v_p^4} e^{-v_0^2/v_p^2} dv_0.
\end{equation}
As a first approximation, we consider a point source at \(x=0\), take \(v_c\) to be independent of \(\theta\), and define the capture angle \(\theta_c\) through geometry, i.e., \(\tan\theta_c = r_a/(z_s-z_a)\). 
The capture velocity then scales as \(v_c\propto B_\text{max}\) and thus the loading rate \(R\propto B_\text{max}^4\).
Fig.~\ref{fig:slower_efficacy} shows the experimental efficacy of our Zeeman slower for four different magnetic field gradients \(B'\).
At most, we observe a factor-of-4 increase in \(R\) for a
doubling of \(B_\text{max}\), suggesting a scaling closer to \(B_\text{max}^2\).

The na\"{\i}ve \(B_\text{max}^4\) scaling breaks down if the acceleration required to keep an atom on the \(v_B(z)\) curve, see Fig.~\ref{fig:slower}, exceeds the maximum possible acceleration from the slowing laser beam. 
This condition is expressed as
\begin{equation}
    \label{eq:slowing_condition}
    \frac{dv_B}{dt} =
    \frac{dv_B}{dz}\frac{dz}{dt} =
    \frac{\mu_B}{\hbar k_0}\frac{dB(z)}{dz} v_B \leq
    \frac{\hbar k_0 \Gamma}{2 m}\frac{s_0}{1+s_0},
\end{equation}
where the right-hand side of the inequality is the maximum magnitude of \(\mathbf{f}\) in the Zeeman slower region (see Eq.~\ref{eq:force}).
In the present study, the largest \(v_B(z_\text{max})\approx 120~\si[per-mode=symbol]{\meter\per\second}\), defined by \(B_\text{max}\approx 13~\si{\milli\tesla}\).
Combining this largest \(v_B(z_\text{max})\) with \(s=3.9\) and the largest \(B'=5.5~\si[per-mode=symbol]{\milli\tesla\per\centi\meter}\) from Fig.~\ref{fig:slower_efficacy}, we find that the inequality in Eq.~\ref{eq:slowing_condition} is always fulfilled.
Because \(dB(z)/dz<B'\) in our apparatus, our calculation also demonstrates that deviations from the ideal \(B(z)\propto\sqrt{z}\) field in the Zeeman slower region cannot explain the observed scaling of \(R\) with \(B_\text{max}\).

Another potential deviation from \(R\propto B_\text{max}^4\) is the assumed independence of \(v_c\) and \(\theta_c\).
Consider the on-axis atomic trajectories in Fig.~\ref{fig:traj} with a nonzero initial angle with the \(z\) axis. 
These trajectories blossom: as the axial velocity decreases, the initial transverse velocity causes the atom to deviate farther off-axis.
The blossoming effect reduces \(\theta_c\).
Neglecting the initial velocity change in the increasing magnetic field region (\(z\gtrsim30~\si{\milli\meter}\) in Fig.~\ref{fig:slower}), an atom will travel at its initial velocity \(v_0\) until it intersects the universal \(v_B(z)\) curve. 
To illustrate the blossoming effect, we take \(v_B(z)\approx v'z\) so that the atom falls onto the universal trajectory at \(z_I = v_0/v'\), where \(v' = \mu_B B'/\hbar k\).
Along the \(v_B(z)\) trajectory, the atom obeys \(z(t) = (v_0/v') e^{-v' t}\).
The angle \(\theta_c\) that just clears the aperture can then be determined from the source-aperture travel time \(t_{sa}\),
\begin{equation}
    \begin{split}
    \tan\theta_c & \approx \frac{r_a}{v_0 t_{sa}} \\
    & = \frac{r_a}{z_s-z_a}\left(\frac{z_s-z_a}{z_s-z_I[1-\log(z_I/z_a)]}\right),
    \end{split}
\end{equation}
where we have assumed that \(\theta_c\ll1\).
For our device, the term in parentheses reduces \(\theta_c\) by approximately \(70~\si{\percent}\) of the geometric limit at \(B' = 5.5~\si[per-mode=symbol]{\milli\tesla\per\centi\meter}\) and \(B_\text{max} = 13~\si{\milli\tesla}\) (see Fig.~\ref{fig:slower_efficacy}).
With this correction, the loading rate, evaluated numerically, scales roughly as \(R\propto B_\text{max}^3\) in our region of experimental interest.

Finally, we consider a source that is misaligned with respect to the slowing laser beam.
Fig.~\ref{fig:traj} shows several such trajectories, which start at \(x\approx5~\si{\milli\meter}\) with various emission angles. 
Simulations of these off-axis trajectories indicate that atoms starting outside the slowing laser beam can still be captured provided that they enter the Zeeman slower at a position \(z_e\) such that \(v_B(z_e)>v_0\) and subsequently clear the aperture.
We use these two conditions to calculate the scaling of the loading rate for each \(B'\), shown in Fig.~\ref{fig:slower_efficacy} as the solid curves.
The model suggests that the vast majority of the flux is being emitted from the source approximately \(2.5~\si{\milli\meter}\) outside of the slowing laser beam, a slight misalignment.

After disassembling the apparatus, we discovered that the Li metal had migrated out of the dispenser onto its exterior surface.
Most of the Li was positioned just outside of the Zeeman slower beam, as suggested by our loading rate calculation.
The Li migration may have been exacerbated because the gravitational force is antiparallel to the \(z\) axis (see Fig.~\ref{fig:sketch}) and might be reduced by reorienting the apparatus.
However, restricting the orientation of the device is undesirable for future applications.
Inserting a Ni mesh into the dispenser to wick the Li metal would prevent migration in all orientations~\cite{Norrgard2018, Gunton2013} and improve the loading rate scaling to \(R\propto B_\text{max}^3\).

The above results suggest that the device would be improved by moving the source further from the MOT while maintaining alignment with the slowing laser beam.
Consider a fixed magnetic field gradient \(B'\) throughout all space and a movable source with position \(z_s\).
The capture velocity for atoms traveling along the \(z\) axis is \(v_c = v' z_s\).
As the source moves away from the MOT, the capture velocity increases because \(v_c \propto z_s\), but \(\theta_c\) decreases as \(\tan\theta_c = r_a/(z_s-z_a)\) for small \(v_0\) and as \(\tan\theta_c\approx r_a v'/[v_0 \log(z_a/z_s)]\) for velocities \(v_0 \lesssim v' z_s\).
To understand the competition between \(v_c\) and \(\theta_c\), we numerically evaluate Eq.~\ref{eq:loading_rate} as a function of \(z_s\).
We find that \(R\) is roughly constant for \(z_s/z_a<3\) and becomes \(R\propto {(z_s/z_a)}^{3/2}\) for \(z_s/z_a>10\).
Therefore, placing the source farther behind the chip is optimal, provided that optical alignment can be maintained, the inequality in Eq.~\ref{eq:slowing_condition} is always satisfied, and the magnetic field gradient can be extended to continue the Zeeman slower.

Loading from behind the chip should always outperform loading from the side, given that \(v_c\) and \(\theta_c\) are roughly equivalent for a source at \(z_s=z_a\) and a source placed to the side (see Fig.~\ref{fig:sketch}).
However, outperforming side loading places additional restrictions on the size of the apparatus (\(z_s/z_a>3\)), beam alignments, and source placement.
As an example, a source placed to the side might have performed equally well in
the device presented here.
We simulated several side-loaded trajectories, shown in green in Fig.~\ref{fig:traj}, and found that the side-loaded capture velocity is roughly \(25~\si{\percent}\) that of the best on-axis capture velocity. 
The reduction in capture velocity is compensated by the increase in \(\theta_c\), causing \(R\) to be unchanged.
Moving our source forward to the position of the peak magnetic field (i.e., \(z_s=z_\text{max}\)) would be sufficient for back loading to be faster than side loading.

\section{Discussion}

We have demonstrated a single-beam slowing and trapping apparatus for \(^7\)Li
atoms.
We trap more than \(10^6\) atoms with loading rates exceeding
\(10^6~\si{\per\second}\).
The integrated Zeeman slower behind the chip is effective in increasing the captured flux by over a factor of three.
The source placement prevents unwanted metal deposition on the grating and allows for future vacuum improvements via differential pumping.

Our design can easily be adapted to serve as cold gas source for a variety of applications.
By implementing differential pumping or using a low-outgassing atom source (rather than a dispenser)~\cite{Kang2017, Barker2018}, our device could be used as a primary vacuum gauge~\cite{Scherschligt2017, Eckel2018}.
The diffraction grating period and etch depth can be altered to optimize trapping of other elements, such as Rb, Cs, Ca, Sr, or Yb.
The trapping of alkaline-earth atoms using our apparatus would allow the development of portable optical frequency standards, which could be used for geodesy~\cite{Chou2010} or space-based gravitational wave detection~\cite{Vutha2015} and will be necessary for future redefinition of the SI second~\cite{Parker2012}.
A multiple-length Z-wire magnetic trap~\cite{Squires2016a} could be patterned onto the back of our grating chip; permitting atoms to be pulled closer to the chip surface for chip-scale atom interferometers~\cite{Fortagh2007, Abend2016} or quantum memories~\cite{Keil2016, Vetsch2010}.
The tetrehedral MOT configuration should also be applicable to ``type-II" MOTs~\cite{Flemming1997, Jarvis2018}, which are used to laser-cool and trap molecules~\cite{Barry2014, Tarbutt2015}.
We anticipate that, with suitable modifications to the grating, our system could trap molecules from a buffer gas beam source~\cite{Hutzler2012}, enabling the development of deployable devices using laser-cooled molecules.

Improvements to the present design can be made to increase both \(R\) and \(N_{S}\).
First, our dispenser could be redesigned to better mode match with our slowing laser beam and could be placed closer to the maximum of the magnetic field.
Together, the changes to the dispenser should increase our loading rate by at least a factor of \(10\).
Second, the Zeeman slower can be shortened by increasing the magnetic field gradient.
Because of the lack of a counterpropagating beam~\cite{Haubrich1993}, the increased magnetic field gradient should not impact the capture velocity, provided that the inequality in Eq.~\ref{eq:slowing_condition} is satisfied.
Third, adding differential pumping, which was not implemented here, will decrease the loss rate \(\tau\) and hence increase the equilibrium number of atoms in the MOT.\@ 

\section*{Acknowledgements}

We thank W. McGehee and S. Maxwell, for their careful reading of the manuscript.
We also thank the CNST NanoFab staff for allowing us to use the facility to fabricate grating chips.
D. S. B. and E. B. N. acknowledge support from the National Research Council Postdoctoral Research Associateship Program.

\bibliography{li_grating_MOT}

\begin{thebibliography}{58}
\bibitem [{dis()}]{disclaimer}%
  \BibitemOpen
  \href@noop {} {}\bibinfo {note} {Commercial fabrication facilities,
  equipment, materials or computational software are identified in this paper
  in order to adequately specify device fabrication, the experimental procedure
  and data analysis. Such identification is not intended to imply endorsement
  by the National Institute of Standards and Technology, nor is it intended to
  imply that the facility, equipment, material or software identified is
  necessarily the best available.}\BibitemShut {Stop}%
\end{thebibliography}

\end{document}